\documentclass[twocolumn,showpacs,amsfonts,aps,prl,floatfix]{revtex4}
\usepackage{amsmath}
\usepackage{bm}
\usepackage{graphicx}
\usepackage{mathrsfs}
\usepackage{footmisc}
\usepackage{multirow}
\usepackage{graphicx}
\usepackage{booktabs, ctable}
\usepackage{xcolor, color, framed}

\begin{document}

\title{Elliptic flow as a probe for $\psi(2S)$ production mechanism in relativistic heavy ion collisions}
\author{Baoyi Chen }
\affiliation{Department of Physics, Tianjin University, Tianjin 300352, China}
\date{\today}

\begin{abstract}
I discuss the elliptic flows of $\psi(2S)$ with different production 
mechanisms in $\sqrt{s_{NN}}=2.76$ TeV Pb-Pb collisions.
If the final $\psi(2S)$s are mainly from the recombination of uncorrelated charm and anticharm quarks at $T\approx T_c$, 
charm and anticharm quarks will carry large collective flows of the bulk medium, which will be inherited by the regenerated $\psi(2S)$s. 
This indicates a larger elliptic flow of $\psi(2S)$ than that of $J/\psi$ which can be regenerated at $T\ge T_c$, 
$v_2^{\psi(2S)}>v_2^{J/\psi}$. However, 
if the final $\psi(2S)$s are mainly from the transitions 
of $J/\psi\rightarrow \psi(2S)$ caused by the color screening of QGP, 
its elliptic flow should be 
close to the elliptic flow of $J/\psi$, $v_2^{\psi(2S)}\sim v_2^{J/\psi}$. 
Therefore, $\psi(2S)$ elliptic flow is a 
sensitive probe for its production mechanisms in relativistic heavy ion collisions. 

\end{abstract}
\maketitle
\section{I. Introduction}
A $J/\psi$ consists of a charm and an anticharm quark with a large binding energy. 
Its abnormal suppression by a deconfined matter has been considered as a signal of the existence of the 
Quark-Gluon Plasma (QGP) produced in heavy ion collisions~\cite{Matsui:1986dk}. 
Charmonium can be dissociated by the color screening~\cite{Satz:2005hx,Zhao:2010nk,Chen:2016dke} and 
the inelastic scatterings~\cite{Peskin:1979va,Bhanot:1979vb,Grandchamp:2001pf,Zhu:2004nw,Thews:2005vj,Chen:2015ona} 
with partons in QGP. Also, the final yields of charmonium can be enhanced by the 
recombination of a charm and an anticharm quark during the evolution of 
QGP~\cite{Thews:2000rj,Zhao:2011cv,Andronic:2003zv,Andronic:2006ky,Yan:2006ve,Chen:2012gg,Chen:2015iga}. 
This mechanism is called the ``regeneration". It even 
dominates the total yield of $J/\psi$ at the available colliding energies of the Large Hadron Collider (LHC)~\cite{Zhao:2011cv,Chen:2015iga}. 
Cold nuclear matter effects, such as shadowing effect~\cite{Eskola:1998df,deFlorian:2003qf,Eskola:2009uj} 
and Cronin effect~\cite{Zhu:2004nw,Zhao:2007hh,Abreu:2000xe,Adare:2006kf}, 
can also change the spatial and momentum distributions of the primordial 
charmonium produced in nucleus-nucleus collisions. 
Different theoretical 
models~\cite{Zhu:2004nw,Thews:2000rj,Andronic:2003zv,Grandchamp:2002wp,Park:2007zza,Song:2010ix,Song:2011kw,Kopeliovich:2014una,Cho:2014xha} 
have been built to explain the experimental data of the 
nuclear modification factor $R_{AA}$, the mean transverse 
momentum squared $\langle p_T^2\rangle$ and the elliptic flow $v_2$ of $J/\psi$. 

Recently, some experimental data of $\psi(2S)$ 
have been published. Different from the ground state $J/\psi$, $\psi(2S)$ is a loosely bound state with a small binding 
energy. Its dissociation temperature is close to the critical temperature of the hadronization transition, 
$T_d(\psi(2S))\approx T_c$~\cite{Satz:2005hx}, which means $\psi(2S)$ $eigenstate$ can barely survive in QGP. 
The CMS Collaboration published the data of prompt $R_{AA}(\psi(2S))\over R_{AA}(J/\psi)$ 
in $\sqrt{s_{NN}}=2.76$ TeV Pb-Pb collisions~\cite{Khachatryan:2014bva}. 
Different mechanisms have been proposed 
for the $\psi(2S)$ prompt and inclusive yields~\cite{Chen:2013wmr,Du:2015wha,Chen:2016vha}. 
These mechanisms include the primordial production at the nucleus colliding time, 
the recombination of a charm and an anticharm quark (or $D$ and $\bar D$ mesons) in 
the later stage of the hot medium evolution and decays from B hadrons. 
Recent studies indicate that the formation time of 
charmonium eigenstates can be delayed by the hot medium in heavy ion collisions~\cite{Song:2015bja}. 
$\psi(2S)$ may suffer less suppression if they are formed later in the gradually cooling QGP. 
With the formation process, 
a $c\bar c$ dipole produced in 
the nucleus-nucleus collisions may exist as a combination of different eigenstates. 
The internal evolution of the $c\bar c$ dipole wavefunction is affected by the hot medium. This changes the fractions of $J/\psi$ and 
$\psi(2S)$ in the $c\bar c$ dipole and the double ratio $R_{AA}(\psi(2S))\over R_{AA}(J/\psi)$~\cite{Chen:2016vha}. 
On the other hand, $J/\psi$ and $\psi(2S)$ can be regenerated at different stage of QGP evolution, and they 
will carry different collective flows of the bulk medium. The elliptic flows $v_2$ of $\psi(2S)$ from coalescence at $T\approx T_c$ 
and transitions of $J/\psi\rightarrow \psi(2S)$ should be different from each other.
Elliptic flow should be a sensitive probe to distinguish which production mechanism dominates the $\psi(2S)$ final yield. 

The article is organized as follows. 
In Sec. II, I introduce the Langevin equation for the charm quark evolution 
and the hydrodynamic equations for the QGP evolution. 
In Sec. III, different mechanisms of the $\psi(2S)$ production are discussed in detail. 
In Sec. IV, I fit the parameters in the Langevin equation to explain the experimental data of D mesons, and then 
give the elliptic flows of charmonium. 
Sec. V is devoted to the summary.

\section{II. Dynamics of heavy quarks in heavy ion collisions}
In this work, I focus on the charmonium regeneration in heavy ion collisions. 
They are mainly from the recombination of charm and anticharm 
quarks in the low transverse momentum bin, where multi-elastic scatterings dominate the energy loss of charm quarks 
~\cite{Braaten:1991we,vanHees:2005wb,Qin:2007rn,Cao:2012au,Cao:2015cba}, and 
the medium-induced gluon radiation~\cite{Wang:1991xy,Gyulassy:1993hr,Baier:1996kr}
is less important. 
In the limit of small momentum transfer, multi quasi-elastic scatterings of 
heavy quarks in QGP can be treated as a Brownian motion and is usually described by the Langevin 
equation~\cite{He:2011qa,Cao:2012jt,Cao:2013ita}, 
\begin{align}
{d\vec p\over dt}=-\eta_D(p)\vec p+\vec \xi
\label{fun-LG}
\end{align}
The first term on the right-hand side are the drag force with the momentum dependence. 
The second term is the random force. Assuming $\vec \xi$ is 
independent of the momentum of each particle, this noise term satisfies the correlation relation:
\begin{align}
\langle \xi^{i}(t)\xi^{j}(t^\prime)\rangle =\kappa \delta ^{ij}\delta(t-t^\prime)
\end{align}
$\kappa$ represents the momentum space diffusion coefficient of heavy quarks. 
The fluctuation-dissipation relation indicates~\cite{Cao:2012jt,Cao:2015cba} 
\begin{align}
\eta_D(p)={\kappa\over 2TE}
\end{align}
$T$ is the temperature of fluid cells in QGP, $E$ is the energy of charm quarks. 
The spatial diffusion coefficient $D$ of heavy quarks is connected with the momentum space diffusion coefficient by 
\begin{align}
D={2T^2\over \kappa}
\end{align}
I follow Ref.\cite{Cao:2015cba} and take $D=C/(2\pi T)$. 
The value of the parameter $C$ can be fixed by 
the experimental data of 
D mesons in $\sqrt{s_{NN}}=2.76$ TeV Pb-Pb collisions.  
 
For numerical evolutions, the Langevin equation can be discretized as~\cite{Cao:2012jt,Cao:2013ita} 
\begin{align}
&\vec p(t+\Delta t) =\vec p(t)-\eta_D(p)\vec p \Delta t+\vec \xi\Delta t\nonumber \\
&\langle \xi^{i}(t)\xi^j(t-n\Delta t)\rangle ={\kappa \over \Delta t}\delta ^{ij}\delta^{0n}
\label{fun-LD}
\end{align}
Here $\Delta t$ is the time step of the numerical evolution. The noise term in Eq.(\ref{fun-LD}) 
is taken to be the Gaussian distribution with the 
width $\sqrt{\kappa/\Delta t}$. 

At the time of nucleus collisions, charm pairs are produced from the parton fusions with a large momentum transfer. 
The number of $c\bar c$ pairs is proportional to the number of binary collisions. 
Without cold nuclear matter effects, 
the spatial distribution of charm quarks 
is proportional to the function  
\begin{align}
{dN_{\rm PbPb}^{c\bar c}\over d\vec x_T}\propto {T_{\rm Pb}(\vec x_T-\vec b/ 2)T_{\rm Pb}(\vec x_T +\vec b/2)\over T_{\rm Pb}(0)T_{\rm Pb}(0)}
\end{align}
Here $T_{\rm Pb}(\vec x_T)=\int dz \rho_{\rm Pb}(\vec x_T, z)$ is the 
thickness function of $\rm Pb$. $\rho_{\rm Pb}(\vec x_T, z)$ is the nucleon density, which is taken to be 
the Woods-Saxon distribution. 
The denominator is for a normalization. 
For the cold nuclear matter effects such as the shadowing effect, 
I employ the EPS09s LO results $r_i^A(x, Q^2,{\vec x}_T)$ 
which already include the spatial dependence in a nucleus~\cite{Helenius:2012wd}. 
Here, $x=(m_T/\sqrt{s_{NN}})\exp(\pm y)$ and $Q^2=m_T^2$~\cite{Vogt:2004dh,Vogt:2010aa}. 
$y$ and $m_T=\sqrt{m_{c\bar c}^2+p_T^2}$ are the rapidity and the 
transverse energy respectively. 
$\vec x_T$ is the transverse coordinate. 
The momentum distribution of charm quarks can be generated by PYTHIA. The shadowing effect is 
included by multiplying charm $p_T$ spectra from PYTHIA by the shadowing factor $r_i^A(x,Q^2,\vec x_T)$. 

The QGP evolutions in heavy ion collisions can be described with (2+1) dimensional ideal hydrodynamics, 
\begin{align}
\label{Eqhydro1}
\partial _\mu T^{\mu\nu}=0 
\end{align}
where $T^{\mu\nu}=(e+p)u^\mu u^\nu-g^{\mu \nu} p$ is the energy-momentum tensor, $u^\mu$ is
the velocity of fluid cells. $e$ and $p$ are the energy density and the pressure. For the equation of state of the medium, 
the deconfined phase is an ideal gas of massless $u$ 
and $d$ quarks, 150 MeV massed $s$ quarks and gluons~\cite{Sollfrank:1996hd}. Hadron
phase is an ideal gas of all known hadrons and resonances
with mass up to 2 GeV~\cite{Hagiwara:2002fs}. With the charged multiplicity at the midrapidity 
$dN_{ch}/dy=1600$~\cite{Gulbrandsen:2013iu,Shen:2012vn}, the maximum 
temperature of QGP at the initial time $\tau_0^{\rm{QGP}}=0.6$ fm/c is initialized to be 
$484$ MeV~\cite{Shen:2012vn}. $\tau_0^{\rm{QGP}}$ is the time of the medium reaching local equilibrium. 

After charm quarks are generated in the spatial and 
the momentum space with cold nuclear matter effects, their evolutions  
in QGP can be simulated by 
the Langevin equation Eq.(\ref{fun-LG}). After the evolutions, one can obtain the nuclear 
modification factor $R_{AA}(p_T)$ and the elliptic flow $v_2(p_T)$ of D mesons.

\section{III. Different mechanisms of the $\psi(2S)$ production}
The dissociation temperature of $\psi(2S)$ eigenstate is close to the critical temperature of deconfined phase transition.  
Sequential regeneration model indicates that the final prompt $\psi(2S)$s are mainly from the 
recombination of uncorrelated charm and anti-charm quarks at the hadronization (and $D$ and $\bar D$ in hadron phase)~\cite{Du:2015wha}. 
In order to calculate $J/\psi$ and $\psi(2S)$ regeneration, 
I employ the Langevin equation for charm quark evolutions in QGP and coalescence model for their recombination at a certain temperature. 
The Wigner function for charm quark recombination is taken as a Gaussian function,
\begin{align}
f(r,q)=A_0 \exp(-r^2/\sigma^2)\exp(-q^2\sigma^2)
\end{align}
$A_0$ is the normalization factor for $\int f(r,q)r^2drq^2dq=1$. 
The Gaussian width is related to the mean-square-radius by 
$\sigma^2=8\langle r_\Psi^2\rangle/3$~\cite{Greco:2003vf}. 
For a charm and an anticharm quarks with a relative distance $r$ and relative momentum $q$, they have a probability 
$P(r,q)=r^2q^2f(r,q)$ to recombine into a charmonium bound state. 
I employ the Monte Carlo method to simulate the coalescence process. If the probability $P(r,q)$ 
is larger than a random number between $0$ and $1$, 
then these charm and anti-charm quarks can recombine into a new charmonium. Considering  
the regenerated charmonium are mainly from the uncorrelated charm pairs, 
charm and anti-charm quarks are generated in nucleus collisions in uncorrelated initial coordinates 
$(\vec x_{c0},\vec q_{c0})$ and $(\vec x_{{\bar c}0}, \vec q_{{\bar c}0})$. 
As this work focus on the effect of QGP collective flows on charmonium production, I neglect the 
the difference between Wigner functions of different charmonium eigenstates, and take 
$\langle r_\Psi^2 \rangle=0.5^2 \rm{\ fm^2}$ from potential model. The additional hot medium suppression on regenerated 
charmonium is also neglected. 

\begin{figure}[!hbt]
\centering
\includegraphics[width=0.42\textwidth]{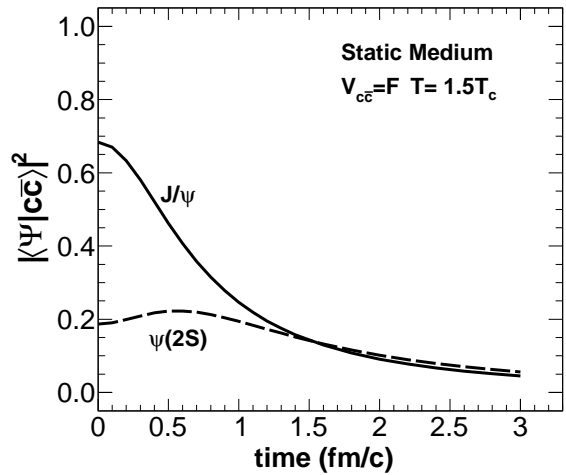}
\caption{(Color online) The time evolution of $J/\psi$ and $\psi(2S)$ fractions in a $c\bar c$ dipole in the static medium 
with a constant temperature $T=1.5T_c$. The initial wavefunction of $c\bar c$ dipole is taken as a 
Gaussian function with the width $\sigma_{0}^{c\bar c}=0.23$ fm.
The heavy quark potential is taken to be the free energy V=F from Lattice results. 
}
\hspace{-0.1mm}
\label{fig-frac}
\end{figure}

The prompt yield of $\psi(2S)$ may also come from correlated $c\bar c$ pairs. 
Correlated $c$ and $\bar c$ are produced with a small 
separation in the spatial space, and need some time to evolve into a certain charmonium 
eigenstate~\cite{Kopeliovich:2014una,Song:2015bja}. 
The dipole with a small size is not likely to be dissociated at the early stage of QGP, which can enhance 
the final production of $J/\psi$ and/or $\psi(2S)$. 
The color screening on heavy quark potential affects the internal evolutions of $c\bar c$ dipoles, which 
corresponds to the transitions between different eigenstates. 
Employing the time-dependent Schr\"odinger equation for the $c\bar c$ dipole internal evolutions in deconfined matter, 
one can evolve the wavefunction of $c\bar c$ dipoles, 
and obtain the fractions of charmonium eigenstates by projecting the $c\bar c$ dipole wavefunction to a certain eigenstate. 
The heavy quark potential at finite temperature is taken to be the free energy $F$ from Lattice results~\cite{Digal:2005ht}, 
The initial wavefunction is taken as a Gaussian function, and the Gaussian width is fitted to satisfy the ratio of 
direct $J/\psi$ and $\psi(2S)$ yields in proton-proton collisions. 
In Fig.\ref{fig-frac}, both fractions of $J/\psi$ and $\psi(2S)$ and the ratio of their 
yields in the $c\bar c$ dipole changes with time. 

Both of the above mechanisms contribute to the $\psi(2S)$ prompt production. 
It would be interesting to find an observable which can 
distinguish the different production mechanisms of $\psi(2S)$. Here, I propose the elliptic flow $v_2$ as a probe 
for the $\psi(2S)$ production. For the final prompt $\psi(2S)$, if most of them are from the regeneration, they 
should be produced at the later stage of the QGP evolution. The elliptic flow of $\psi(2S)$ will be much larger than the 
elliptic flow of $J/\psi$, see Fig.\ref{fig-psiv2} in Section IV. However, if most of the prompt $\psi(2S)$ are from 
the correlated $c\bar c$ dipoles with the formation process, 
then the elliptic flow of $\psi(2S)$ 
$v_2^{\psi(2S)}(p_T)$ should be similar to $v_2^{J/\psi}(p_T)$. 
The detailed discussions are given in Section IV.

\section{IV. Observables of the charm flavor} 
\begin{figure}[!hbt]
\centering
\includegraphics[width=0.42\textwidth]{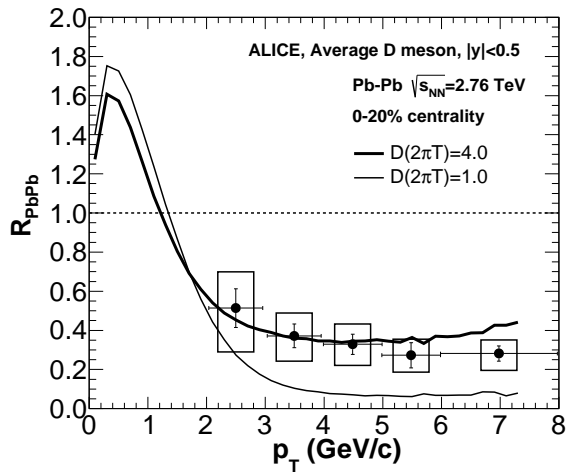}
\caption{The nuclear modification factor $R_{AA}$ of D mesons as a function of the transverse momentum $p_T$ in $\sqrt{s_{NN}}=2.76$ TeV 
Pb-Pb collisions with different diffusion coefficients. 
The thick and thin solid lines correspond to the situations of $C=4.0$ and $C=1.0$ 
respectively. The experimental data is from the ALICE Collaboration~\cite{ALICE:2012ab}.  }
\hspace{-0.1mm}
\label{fig-RAA-pt}
\end{figure}

\begin{figure}[!hbt]
\centering
\includegraphics[width=0.42\textwidth]{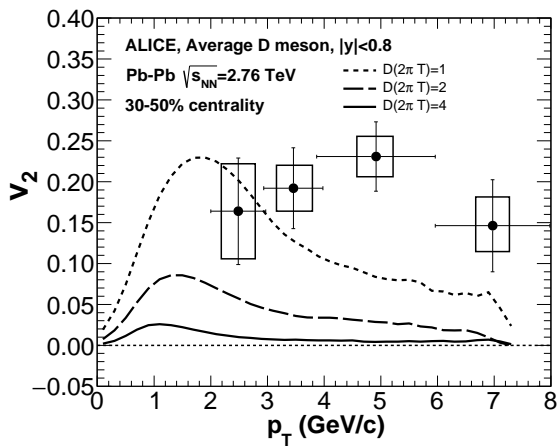}
\caption{The elliptic flow of D mesons as a function of the transverse momentum $p_T$ in $\sqrt{s_{NN}}=2.76$ TeV 
Pb-Pb collisions with different diffusion coefficients. The solid, dashed and dotted lines correspond to the situation of $C=4,2,1$ 
respectively. The experimental data is from the ALICE Collaboration~\cite{Abelev:2013lca}. }
\hspace{-0.1mm}
\label{fig-Dv2}
\end{figure}

The evolutions of heavy quarks in the hot medium can be described by the Langevin equation. 
Different drag coefficients are employed in different models~\cite{Das:2015ana}.
I fit the experimental data of D mesons with different values of the parameter $C$, see Fig.\ref{fig-RAA-pt}-\ref{fig-Dv2}. 
At the critical temperature $T_c$, the deconfined matter is transformed into the hadron gas. 
Charm quarks are transformed to D mesons with coalescence and fragmentation~\cite{Cao:2013ita,Ravagli:2007xx,He:2010vw}. 
The process of hadronization can 
shift the $v_2(p_T)$ by about a 20-25\% upward~\cite{Fries:2011dj}. 
Both collective flows of the bulk medium and D mesons are mainly developed in the deconfined phase. 
In this work, my intent is to employ a reasonable drag coefficient inspired by the 
experimental data of D mesons, and show the big difference between $\psi(2S)$ elliptic flows 
with different production mechanisms. 
Therefore, I neglect the process of charm quarks transforming to D mesons 
and the evolutions of D mesons in hadron gas. These simplifications 
should not change the conclusions about the elliptic flows of charmonium in Section IV and V.

\begin{figure}[!hbt]
\centering
\includegraphics[width=0.42\textwidth]{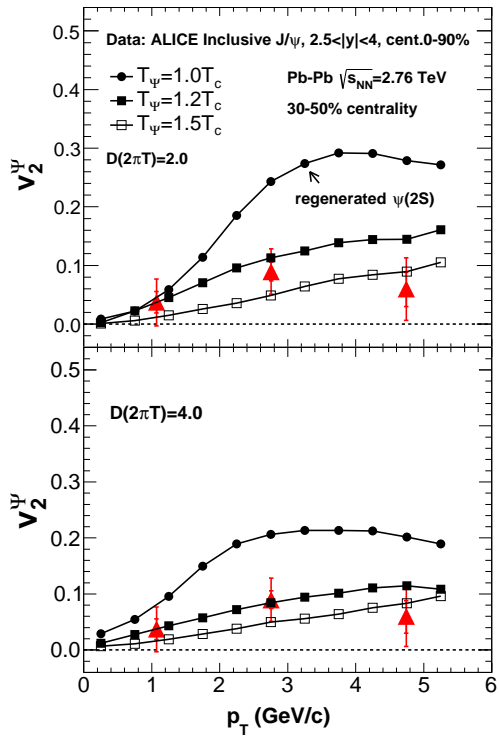}
\caption{(Color online) The elliptic flow of $\Psi=(J/\psi,\psi(2S))$ as a 
function of the transverse momentum $p_T$ in $\sqrt{s_{NN}}=2.76$ TeV 
Pb-Pb collisions. The solid-circle, solid-square and hollow-square lines 
correspond to the situations that 
$\Psi$ are regenerated from the recombination of charm and anticharm quarks at $T_\Psi=(1.5,1.2,1.0)T_c$ 
respectively. Data is from the ALICE Collaboration~\cite{Yang:2012fw}. }
\hspace{-0.1mm}
\label{fig-psiv2}
\end{figure}

When heavy quarks move in the QGP, they lose energy and carry collective flows of the bulk medium. 
It seems difficult to explain $R_{AA}(p_T)$ and $v_2(p_T)$ of D mesons at the 
same time at $\sqrt{s_{NN}}=2.76$ TeV Pb-Pb collisions~\cite{Das:2015ana}.
With smaller value of the parameter $C$, heavy quarks are easier to be thermalized in the QGP. 
This will result in a stronger suppression of $R_{AA}(p_T)$ in the high $p_T$ bin (thin line in Fig.\ref{fig-RAA-pt}) 
and a stronger elliptic flow 
of charm quarks (dotted line in Fig.\ref{fig-Dv2}). In Fig.\ref{fig-Dv2}, lines and data points are for 
the charm quark and D meson elliptic flows respectively. 
Considering the additional 
hadronization process and the hadron phase effects will shift the lines upward, 
the value of $C=2$ is employed for the prediction of $\psi(2S)$ elliptic flows in Fig.\ref{fig-psiv2}. 

The final prompt charmonium consists of three parts: primordial production at the nucleus colliding time, 
the regeneration from the recombination of $c$ and $\bar c$ (or $D$ and $\bar D$) 
during the evolution of the hot medium, and the transitions 
from other charmonium eigenstates. With a realistic description of charm quark evolution, one can 
obtain the distributions of the regenerated charmonium. 
Compared with $J/\psi$, the regeneration of $\psi(2S)$ can only happen 
at the later stage of the 
QGP evolution due to its small binding energy. At that time, the collective flows of QGP 
are stronger. Therefore, the elliptic flow of $\psi(2S)$ should be 
much larger than the elliptic flow of $J/\psi$ which can be regenerated in a relatively earlier time of the QGP evolution.

After charmonium is produced, 
their elliptic flows are almost not changed anymore. 
(for example, the elliptic flow of the primordially 
produced $J/\psi$ is close to zero)~\cite{Zhou:2014kka}. 
As the binding energy of $J/\psi$ is large, they can be regenerated at $T\ge T_c$. In Fig.\ref{fig-psiv2}, 
let's assume that a certain eigenstate $\Psi$ is 
regenerated at $(1.5, 1.2, 1.0)T_c$ respectively. Its elliptic flow can be obtained 
(see solid-circle, solid-square and hollow-square lines in 
Fig.\ref{fig-psiv2}). The dissociation temperature of $J/\psi$ is 
around $T_d^{J/\psi}=(1.5-2.0)T_c$~\cite{Satz:2005hx}. Therefore the regeneration of 
$J/\psi$ happens during $T_c\le T^{\rm{QGP}}<T_d^{J/\psi}$. The elliptic flow 
of the situation $T_\Psi=1.2T_c$ is close to the experimental data 
of inclusive $J/\psi$~\cite{Yang:2012fw}. For the final prompt $\psi(2S)$s, if they are from 
the transitions of $J/\psi$, $v_2^{\psi(2S)}(p_T)$ should be close to 
the dashed line. If the final prompt $\psi(2S)$s are mainly from the regeneration, 
they should be regenerated 
at $T^{\rm{QGP}}\approx T_c$. And the elliptic flow of $\psi(2S)$ should be close to the dotted line. 
Different mechanisms result in very 
different elliptic flows of $\psi(2S)$, 
which makes $v_2^{\psi(2S)}(p_T)$ a sensitive probe for the $\psi(2S)$ production mechanism.

The elliptic flows in Fig.\ref{fig-psiv2} only include the regenerated charmonium. After including 
the primordially produced charmonium, 
the lines at $p_T>3$ GeV/c will be shifted downward a little and approach zero at very high $p_T$ bin. 
But it does not change the relation between three lines in Fig.\ref{fig-psiv2}.
With different forms of the drag coefficient, as long as the regeneration dominates the final 
yield, the elliptic flow of $\psi(2S)$ should be larger than that of $J/\psi$. In the other situation, 
they should be similar to each other. If the drag coefficient is larger at a lower 
temperature (see the parametrization in Ref.\cite{Das:2015ana}), the difference between elliptic flows of 
regenerated $J/\psi$ and $\psi(2S)$ will be even larger. 
In a more realistic situation, 
charmonium should be regenerated in a temperature region, not at a certain temperature $T_\Psi$. 
Different choices of heavy quark potential at finite temperature also affect the regeneration process. 
Both of these effects can be approximated by employing 
different values of $T_\Psi$ in Fig.\ref{fig-psiv2}. 
These will be treated more seriously in the future works.

\section{V. conclusion}
In summary, I employ the Langevin equation to describe the charm quark evolutions and Wigner function for charmonium 
regeneration in QGP. Different production mechanisms are discussed for the $\psi(2S)$ prompt production 
in $\sqrt{s_{NN}}=2.76$ TeV Pb-Pb collisions. 
The elliptic flow of $\psi(2S)$ is proposed as a sensitive probe to distinguish the different production mechanisms. 
If the final prompt $\psi(2S)$ are mainly from the $correlated$ $c\bar c$ dipoles, 
the elliptic flow of $\psi(2S)$ $v_2^{\psi(2S)}(p_T)$ should be close to $v_2^{J/\psi}(p_T)$. 
The prompt $\psi(2S)$ may also come from the 
recombination of $uncorrelated$ $c$ and $\bar c$, which happens at the later 
stage of the QGP evolution. 
In this situation, charm quarks 
carry large collective flows, which will be inherited by the regenerated $\psi(2S)$s. 
Therefore the elliptic flow of $\psi(2S)$ is much larger than that of $J/\psi$. 
The relation between $v_2^{\psi(2S)}(p_T)$ and $v_2^{J/\psi}(p_T)$ is sensitive to the production mechanisms of $\psi(2S)$. 
With different drag coefficients in the Langevin equation, the conclusions about 
the relations between $J/\psi$ and $\psi(2S)$ elliptic flows do not change. This makes the elliptic flow 
a sensitive and robust probe for the $\psi(2S)$ production mechanism. 

\vspace{1.0cm}
{\bf Acknowledgement: } Author thanks Dr. Y. Liu, M. He, P. Zhuang and R. Rapp for helpful discussions, 
and J. Zhao for proof reading the manuscript. 
The work is supported by the NSFC under Grant No. 11547043.


\begin{thebibliography}{20}
\bibitem{Matsui:1986dk} 
  T.~Matsui and H.~Satz,
  Phys.\ Lett.\ B {\bf 178}, 416 (1986).
\bibitem{Satz:2005hx} 
  H.~Satz,
  J.\ Phys.\ G {\bf 32}, R25 (2006)
\bibitem{Zhao:2010nk} 
  X.~Zhao and R.~Rapp,
  Phys.\ Rev.\ C {\bf 82}, 064905 (2010)

\bibitem{Chen:2016dke} 
  B.~Chen, T.~Guo, Y.~Liu and P.~Zhuang,
  arXiv:1607.07927 [nucl-th].

\bibitem{Peskin:1979va} 
  M.~E.~Peskin,
  Nucl.\ Phys.\ B {\bf 156}, 365 (1979).
\bibitem{Bhanot:1979vb} 
  G.~Bhanot and M.~E.~Peskin,
  Nucl.\ Phys.\ B {\bf 156}, 391 (1979).

\bibitem{Grandchamp:2001pf} 
  L.~Grandchamp and R.~Rapp,
  Phys.\ Lett.\ B {\bf 523}, 60 (2001)
\bibitem{Zhu:2004nw} 
  X.~l.~Zhu, P.~f.~Zhuang and N.~Xu,
  Phys.\ Lett.\ B {\bf 607}, 107 (2005)
\bibitem{Thews:2005vj} 
  R.~L.~Thews and M.~L.~Mangano,
  Phys.\ Rev.\ C {\bf 73}, 014904 (2006)

\bibitem{Chen:2015ona} 
  B.~Chen, 
  Phys.\ Rev.\ C {\bf 93}, no. 4, 044917 (2016)

\bibitem{Thews:2000rj} 
  R.~L.~Thews, M.~Schroedter and J.~Rafelski,
  Phys.\ Rev.\ C {\bf 63}, 054905 (2001)
\bibitem{Zhao:2011cv} 
  X.~Zhao and R.~Rapp,
  Nucl.\ Phys.\ A {\bf 859}, 114 (2011)
\bibitem{Andronic:2003zv} 
  A.~Andronic, P.~Braun-Munzinger, K.~Redlich and J.~Stachel,
  Phys.\ Lett.\ B {\bf 571}, 36 (2003)
\bibitem{Andronic:2006ky} 
  A.~Andronic, P.~Braun-Munzinger, K.~Redlich and J.~Stachel,
  Nucl.\ Phys.\ A {\bf 789}, 334 (2007)
\bibitem{Yan:2006ve} 
  L.~Yan, P.~Zhuang and N.~Xu,
  Phys.\ Rev.\ Lett.\  {\bf 97}, 232301 (2006)
\bibitem{Chen:2012gg} 
  B.~Chen, K.~Zhou and P.~Zhuang,
  Phys.\ Rev.\ C {\bf 86}, 034906 (2012)

\bibitem{Chen:2015iga} 
  B.~Chen,
  Phys.Rev.C.93.054905(2016)

\bibitem{Eskola:1998df} 
  K.~J.~Eskola, V.~J.~Kolhinen and C.~A.~Salgado,
  Eur.\ Phys.\ J.\ C {\bf 9}, 61 (1999)
\bibitem{deFlorian:2003qf} 
  D.~de Florian and R.~Sassot,
  Phys.\ Rev.\ D {\bf 69}, 074028 (2004)

\bibitem{Eskola:2009uj} 
  K.~J.~Eskola, H.~Paukkunen and C.~A.~Salgado,
  JHEP {\bf 0904}, 065 (2009)
\bibitem{Zhao:2007hh} 
  X.~Zhao and R.~Rapp,
  Phys.\ Lett.\ B {\bf 664}, 253 (2008)
\bibitem{Abreu:2000xe} 
  M.~C.~Abreu {\it et al.} [NA50 Collaboration],
  Phys.\ Lett.\ B {\bf 499}, 85 (2001).
\bibitem{Adare:2006kf} 
  A.~Adare {\it et al.} [PHENIX Collaboration],
  Phys.\ Rev.\ Lett.\  {\bf 98}, 232002 (2007)

\bibitem{Grandchamp:2002wp} 
  L.~Grandchamp and R.~Rapp,
  Nucl.\ Phys.\ A {\bf 709}, 415 (2002)
\bibitem{Park:2007zza} 
  Y.~Park, K.~I.~Kim, T.~Song, S.~H.~Lee and C.~Y.~Wong,
  Phys.\ Rev.\ C {\bf 76}, 044907 (2007)
\bibitem{Song:2010ix} 
  T.~Song, W.~Park and S.~H.~Lee,
  Phys.\ Rev.\ C {\bf 81}, 034914 (2010)
\bibitem{Song:2011kw} 
  T.~Song, W.~Park and S.~H.~Lee,
  Phys.\ Rev.\ C {\bf 84}, 054903 (2011)
\bibitem{Kopeliovich:2014una} 
  B.~Z.~Kopeliovich, I.~K.~Potashnikova, I.~Schmidt and M.~Siddikov,
  Phys.\ Rev.\ C {\bf 91}, no. 2, 024911 (2015)

\bibitem{Cho:2014xha} 
  S.~Cho,
  Phys.\ Rev.\ C {\bf 91}, no. 5, 054914 (2015)


\bibitem{Khachatryan:2014bva} 
  V.~Khachatryan {\it et al.} [CMS Collaboration],
  Phys.\ Rev.\ Lett.\  {\bf 113}, no. 26, 262301 (2014)

\bibitem{Chen:2013wmr} 
  B.~Chen, Y.~Liu, K.~Zhou and P.~Zhuang,
  Phys.\ Lett.\ B {\bf 726}, 725 (2013)

\bibitem{Du:2015wha} 
  X.~Du and R.~Rapp,
  Nucl.\ Phys.\ A {\bf 943}, 147 (2015)
\bibitem{Chen:2016vha} 
  B.~Chen, X.~Du and R.~Rapp,
  arXiv:1612.02089 [nucl-th]

\bibitem{Song:2015bja} 
  T.~Song, C.~M.~Ko and S.~H.~Lee,
  Phys.\ Rev.\ C {\bf 91}, no. 4, 044909 (2015)

\bibitem{Braaten:1991we} 
  E.~Braaten and M.~H.~Thoma,
  Phys.\ Rev.\ D {\bf 44}, 2625 (1991).
  %
\bibitem{vanHees:2005wb} 
  H.~van Hees, V.~Greco and R.~Rapp,
  Phys.\ Rev.\ C {\bf 73}, 034913 (2006)

\bibitem{Qin:2007rn} 
  G.~Y.~Qin, J.~Ruppert, C.~Gale, S.~Jeon, G.~D.~Moore and M.~G.~Mustafa,
  Phys.\ Rev.\ Lett.\  {\bf 100}, 072301 (2008)
\bibitem{Cao:2012au} 
  S.~Cao, G.~Y.~Qin, S.~A.~Bass and B.~Müller,
  Nucl.\ Phys.\ A {\bf 904-905}, 653c (2013)
\bibitem{Cao:2015cba} 
  S.~Cao, G.~Y.~Qin and S.~A.~Bass,
  Phys.\ Rev.\ C {\bf 92}, no. 5, 054909 (2015)

\bibitem{Wang:1991xy} 
  X.~N.~Wang and M.~Gyulassy,
  Phys.\ Rev.\ Lett.\  {\bf 68}, 1480 (1992).
\bibitem{Gyulassy:1993hr} 
  M.~Gyulassy and X.~n.~Wang,
  Nucl.\ Phys.\ B {\bf 420}, 583 (1994)
\bibitem{Baier:1996kr} 
  R.~Baier, Y.~L.~Dokshitzer, A.~H.~Mueller, S.~Peigne and D.~Schiff,
  Nucl.\ Phys.\ B {\bf 483}, 291 (1997)

\bibitem{He:2011qa} 
  M.~He, R.~J.~Fries and R.~Rapp,
  Phys.\ Rev.\ C {\bf 86}, 014903 (2012)

\bibitem{Cao:2012jt} 
  S.~Cao, G.~Y.~Qin and S.~A.~Bass,
  J.\ Phys.\ G {\bf 40}, 085103 (2013)
\bibitem{Cao:2013ita} 
  S.~Cao, G.~Y.~Qin and S.~A.~Bass,
  Phys.\ Rev.\ C {\bf 88}, 044907 (2013)

\bibitem{Helenius:2012wd} 
  I.~Helenius, K.~J.~Eskola, H.~Honkanen and C.~A.~Salgado,
  JHEP {\bf 1207}, 073 (2012)


\bibitem{Vogt:2004dh} 
  R.~Vogt,
  Phys.\ Rev.\ C {\bf 71}, 054902 (2005)
\bibitem{Vogt:2010aa} 
  R.~Vogt,
  Phys.\ Rev.\ C {\bf 81}, 044903 (2010)

\bibitem{Sollfrank:1996hd} 
  J.~Sollfrank, P.~Huovinen, M.~Kataja, P.~V.~Ruuskanen, M.~Prakash and R.~Venugopalan,
  Phys.\ Rev.\ C {\bf 55}, 392 (1997)

\bibitem{Hagiwara:2002fs} 
  K.~Hagiwara {\it et al.} [Particle Data Group Collaboration],
  Phys.\ Rev.\ D {\bf 66}, 010001 (2002).

\bibitem{Gulbrandsen:2013iu} 
  K.~Gulbrandsen [ALICE Collaboration],
  J.\ Phys.\ Conf.\ Ser.\  {\bf 446}, 012027 (2013)
\bibitem{Shen:2012vn} 
  C.~Shen and U.~Heinz,
  Phys.\ Rev.\ C {\bf 85}, 054902 (2012)
  Erratum: [Phys.\ Rev.\ C {\bf 86}, 049903(E) (2012)]

\bibitem{Greco:2003vf} 
  V.~Greco, C.~M.~Ko and R.~Rapp,
  Phys.\ Lett.\ B {\bf 595}, 202 (2004)


%
\bibitem{Digal:2005ht} 
  S.~Digal, O.~Kaczmarek, F.~Karsch and H.~Satz,
  Eur.\ Phys.\ J.\ C {\bf 43}, 71 (2005)

\bibitem{Das:2015ana} 
  S.~K.~Das, F.~Scardina, S.~Plumari and V.~Greco,
  Phys.\ Lett.\ B {\bf 747}, 260 (2015)

\bibitem{Ravagli:2007xx} 
  L.~Ravagli and R.~Rapp,
  Phys.\ Lett.\ B {\bf 655}, 126 (2007)
\bibitem{He:2010vw} 
  M.~He, R.~J.~Fries and R.~Rapp,
  Phys.\ Rev.\ C {\bf 82}, 034907 (2010)

\bibitem{Fries:2011dj} 
  R.~J.~Fries, M.~He and R.~Rapp,
  J.\ Phys.\ G {\bf 38}, 124068 (2011)


\bibitem{ALICE:2012ab} 
  B.~Abelev {\it et al.} [ALICE Collaboration],
  JHEP {\bf 1209}, 112 (2012)

\bibitem{Abelev:2013lca} 
  B.~Abelev {\it et al.} [ALICE Collaboration],
  Phys.\ Rev.\ Lett.\  {\bf 111}, 102301 (2013)

\bibitem{Zhou:2014kka} 
  K.~Zhou, N.~Xu, Z.~Xu and P.~Zhuang,
  Phys.\ Rev.\ C {\bf 89}, no. 5, 054911 (2014)


\bibitem{Yang:2012fw} 
  H.~Yang [ALICE Collaboration],
  Nucl.\ Phys.\ A {\bf 904-905}, 673c (2013)





\end{thebibliography}
\end{document}